\documentclass[12pt]{article}
\usepackage{amssymb}
\usepackage{tikz}
\usetikzlibrary{arrows, calc, decorations.pathmorphing, backgrounds, positioning, fit, petri, automata}
\definecolor{grey1}{rgb}{0.5,0.5,0.5}
\usepackage[top=2.54cm, bottom=2.54cm, left=2.2cm, right=2.2cm]{geometry}

\usepackage{graphicx}
\usepackage{subfigure}
\usepackage{amsmath,amsthm,bm}
\usepackage{rotating}
\usepackage{chngcntr}
\usepackage{apptools}
\usepackage[titletoc,title]{appendix}
\usepackage{comment}

\usepackage{hyperref}
\usepackage{xcolor}         
\definecolor{grau}{rgb}{0.8,0.8,0.8}

\newcommand{\chen}[1]{\color{orange}}
\usepackage{setspace,lscape,longtable}
\usepackage{amsmath,amsthm,bm}
\usepackage{fullpage}  

\usepackage{titlesec}
\usepackage[english]{babel}
\newtheoremstyle{mystyle}{1.5ex plus 1ex minus .2ex}{1.5ex plus 1ex minus.2ex}{}{}{\bf}{}{1em}{}
\theoremstyle{mystyle}

\usepackage[comma,authoryear]{natbib}

\bibpunct{(}{)}{;}{a}{,}{,}
\numberwithin{equation}{section}

\newcommand{\bY}{\bm{Y}}

\newcommand{\by}{\bm{y}}

\newcommand{\balpha}{\bm{\alpha}}

\newcommand{\btheta}{\bm{\theta}}

\newcommand{\bphi}{\bm{\phi}}

\newcommand{\Poi}{\text{Pois}}

\DeclareMathOperator*{\argmax}{argmax}

\newcommand{\cdotv}{\boldsymbol \cdot}

\newcommand{\zz}{\color{orange} \it}

\newcommand{\xx}{\color{black}\rm  }    

\author{Fangzheng Xie\thanks{Department of Applied Mathematics and Statistics, Johns Hopkins University, Baltimore, MD 21218, USA. } \and Mingyuan Zhou\thanks{Department of Information, Risk, \& Operations Management and Department of Statistics \& Data Sciences, The University of Texas at Austin,  Austin, TX 78712, USA.}
	 \and Yanxun Xu\footnotemark[1] \thanks{Correspondence should be addressed to Yanxun Xu (yanxun.xu@jhu.edu)}}
\title{BayCount: A Bayesian Decomposition 
Method for Inferring  Tumor  Heterogeneity using RNA-Seq Counts}
\date{}

\begin{document}
\doublespacing

\maketitle

\begin{abstract}
Tumor is heterogeneous -- a tumor sample usually consists of a set of subclones with distinct transcriptional profiles and potentially different degrees of aggressiveness and responses to drugs. Understanding tumor heterogeneity is therefore critical to precise cancer prognosis and treatment. 
In this paper, we introduce BayCount, a Bayesian decomposition method to infer tumor heterogeneity with highly over-dispersed  RNA sequencing count data. 
Using negative binomial factor analysis, 
BayCount takes into account both the between-sample  and gene-specific random effects 
on raw counts of sequencing reads mapped to each gene.
For posterior inference, we develop an efficient compound Poisson based 
blocked Gibbs sampler.  
Through extensive simulation studies and analysis of The Cancer Genome Atlas lung cancer and kidney cancer RNA sequencing count data, we  show that 
BayCount is able to accurately estimate the number of subclones,  the proportions of these subclones in each tumor sample, and the gene expression profiles in each subclone. Our method represents the first effort in characterizing tumor heterogeneity using RNA sequencing 
 count data that simultaneously removes the need of normalizing the counts, 
achieves statistical robustness, and obtains biologically/clinically meaningful insights. 

\end{abstract}

\section{Introduction}
\label{sec:intro}
Tumor heterogeneity (TH) is a phenomenon that describes distinct molecular profiles of different cells in one or more tumor samples. 
TH arises during the formation of a tumor as a fraction of cells acquire and accumulate different somatic events (\emph{e.g.}, mutations in different cancer genes), resulting in heterogeneity within the same biological tissue sample and between different ones,
spatially and temporally \citep{russnes2011insight, ding2012clonal}. As a result, tumor cell populations are composed of different subclones (subpopulations) of cells, characterized by distinct genomes, transcriptional profiles \citep{kim2015single}, as well as other molecular profiles, such as copy number alterations.  
Understanding TH is critical to precise cancer prognosis and treatment. Heterogenetic tumors may exhibit different degrees of aggressiveness and responses  to drugs among different samples due to genetic or gene expression differences. The level of heterogeneity itself can be used as a biomarker to predict treatment response or prognosis since more heterogeneous tumors are more likely to contain treatment-resistant subclones \citep{marusyk2012intra}. 
This will ultimately facilitate the rational design of combination treatments, with each distinct compound targeting a specific tumor subclone based on its transcriptional profile.

Large-scale sequencing techniques provide  valuable information for understanding tumor complexity and open a door for the desired statistical inference on TH. Previous studies have focused on reconstructing the subclonal composition by quantifying  the  structural subclonal copy number variations  \citep{carter2012absolute,oesper2013theta},  somatic mutations   
\citep{nik2012life,roth2014pyclone,xu2015mad}, or both \citep{deshwar2015phylowgs, lee2016bayesian}. In this paper, we aim to learn tumor transcriptional heterogeneity using RNA sequencing (RNA-Seq) data. 

In the analysis of  gene expression data, matrix decomposition models have been extensively studied in the context of microarray and {\it normalized} RNA-Seq data \citep{venet2001separation,lahdesmaki2005silico,wang2006computational,abbas2009deconvolution, repsilber2010biomarker,shen2010cell, gong2011optimal, hore2016tensor,wang2016mathematical}. Generally, given gene expression data matrix $X=(x_{ij})_{G\times S}$, where the $(i,j)$th element records the expression value of the $i$th gene in the $j$th sample, they decompose $X$ by modeling  $x_{ij}$ with $\sum_{k=1}^K\phi_{ik}\theta_{kj}$,
where $\phi_{ik}$ encodes the expression level of the $i$th gene in the $k$th subclone, $\theta_{kj}$ represents the mixing 
weight of the $k$th subclone in the $j$th sample, and $K$ is the number of subclones. The decomposition can be solved by either optimization algorithms \citep{venet2001separation, wang2016mathematical} or statistical inference by assuming a normal distribution on~$x_{ij}$. While it is reasonable to assume normality for microarray gene expression data, it is often inappropriate to adopt such an assumption for directly modeling RNA-Seq data, which involve nonnegative integer observations. 
If 
a model based on normal distribution is used,  one often needs to first normalize RNA-Seq data 
 before performing any downstream analysis.  See \citet{dillies2013comprehensive} for a review on normalization methods. 
 Although normalization often destroys the nonnegative and discrete nature of the RNA-Seq data, it remains the predominant way for data preprocessing 
 due to not only the computational convenience in modeling normalized data, but also the lack of appropriate count data models. Distinct from previously proposed methods, in this paper, we  propose an attractive class of count data models in 
 decomposing RNA-Seq count matrices.

There are, nevertheless, statistical challenges with RNA-Seq count data. First, the distributions of the RNA-Seq count data 
are typically over-dispersed and  sparse. Second, the scales of the read counts in sequencing data across samples can be enormously different  due to the mechanism of the sequencing experiment such as the variations in technical lane capacities. The larger the library sizes (\emph{i.e.}, sequencing depth) are, the larger the read counts tend to be. In addition, the differences in gene lengths or GC-content \citep{pickrell2010understanding} can bias 
gene differential expression analysis, particularly for lowly expressed genes \citep{oshlack2009transcript}. A number of count data models have been developed for RNA-Seq data \citep{lee2013poisson, kharchenko2014bayesian, fan2016characterizing}. For example, 
\cite{lee2013poisson}  proposed a Poisson factor model on microRNA to reduce the dimension of count data and identify low-dimensional features, followed by a clustering procedure over tumor samples. \cite{kharchenko2014bayesian} developed a method using a mixture of negative binomial and Poisson distributions to model single cell RNA-Seq data for gene differential expression  analysis. None of these methods, however, address the problem of TH. 

To this end, we propose BayCount, a Bayesian 
matrix decomposition model 
built upon the negative binomial model  \citep{zhou2016nonparametric}, to infer tumor transcriptional heterogeneity using RNA-Seq count data. BayCount accounts for 
both the  between-sample and gene-specific random effects and infers the number of latent subclones, the proportions of these subclones in each sample, and subclonal expression simultaneously.  

The remainder of the paper is organized as follows. 
 In Section \ref{sec:probability_model}, we  introduce BayCount, a hierarchical Bayesian model for RNA-Seq  count data, and develop an  efficient compound Poisson based blocked Gibbs sampler. We investigate the performance of posterior inference and robustness of the BayCount model through extensive simulation studies in Section 3, and apply our proposed BayCount model to analyze two 
 real-world RNA-Seq datasets from The Cancer Genome Atlas (TCGA) \citep{cancer2012comprehensive}
 in Section 4. We conclude the paper 
 in Section \ref{sec:con}.


\section{Hierarchical Bayesian Model and Inference}\label{sec:probability_model}
In this section we  present the proposed hierarchical model for RNA-Seq count data, develop the corresponding posterior inference, and discuss how to 
determine the number of subclones. 

\subsection{BayCount Model}
We assume that $S$ 
 tumor samples are available from the same or different patients. 
Consider a $G\times S$ count matrix $Y=(y_{ij})_{G\times S}$, where each row represents a gene, each column represents a tumor sample, and the element $y_{ij}$ records the read count of the $i$th gene from the $j$th tumor sample. The Poisson distribution $\mathrm{Pois}(\lambda)$ with mean $\lambda>0$ is commonly used for modeling count data.
Poisson factor analysis (PFA) \citep{zhou2012beta}  factorizes the count matrix $Y$ as 
$y_{ij}\sim\Poi\left(\sum_{k=1}^K\phi_{ik}\theta_{kj}\right)$, 
where $\Phi=(\phi_{ik})_{G\times K}\in\mathbb{R}_+^{G\times K}$ is 
the factor loading matrix 
and $\Theta=(\theta_{kj})_{K\times S}\in\mathbb{R}_+^{K\times S}$ is the factor score matrix. 
Here $K$ is an integer indicating the number of latent factors, and each column of $\Phi$ is subject to the constraint that $\sum_{i=1}^G\phi_{ik}=1$ and $\phi_{ij}\ge 0$. However, the restrictive equidispersion property of the Poisson distribution that the variance and mean are the same limits the application of PFA in modeling sequencing data, which are often highly over-dispersed.
For this reason, one may consider 
negative binomial factor analysis (NBFA)  of \citet{zhou2016nonparametric} that factorizes $Y$ as
$y_{ij}\sim\mathrm{NB}\left(\sum_{k=1}^K\phi_{ik}\theta_{kj},p_j\right)$, where $p_j\in(0, 1)$. 
We denote $y\sim\mathrm{NB}(r, p)$ as a negative binomial distribution with shape parameter $r>0$ and success probability $p\in(0,1)$, whose 
mean and variance 
are $rp/(1-p)$ and $rp/(1-p)^2$, respectively, with the variance-to-mean ratio as $1/(1-p)$.

Denote the $j$th column of $\bY$ as $\by_j=(y_{1j}, y_{2j}, \dots, y_{Gj})^T$,  the count profile of the $j$th tumor sample. 
To account for both the between-sample and gene-specific random effects when modeling RNA-Seq count data, we  propose 
\begin{eqnarray}\label{NBFA_for_TH}
y_{ij}\mid \lambda,\alpha_i,\zeta_j, p_j,\Phi,\Theta &\sim&\mathrm{NB}\left(\lambda\alpha_i+\sum_{k=1}^K\phi_{ik}\theta_{kj}\zeta_j,~p_j\right),
\end{eqnarray}
where $\alpha_i$ accounts for the gene-specific random effect of the $i$th gene, $\lambda$ and $p_j$ control the scales of the gene-specific effect and between-sample effect of the $j$th sample, respectively,  and $\sum_{k=1}^K\phi_{ik}\theta_{kj}\zeta_j$ represents the average effect of the $K$ subclones on the expression of the $i$th gene in the $j$th sample.

To see this, recall that the mean of $y_{ij}$
based on \eqref{NBFA_for_TH} is
\begin{eqnarray}\label{NBFA_e}
\mathbb{E}[y_{ij}]=\left(\lambda\alpha_i+\sum_{k=1}^K\phi_{ik}\theta_{kj}\zeta_j\right) \frac{p_j}{1-p_j}.
\end{eqnarray}
Since $p_j$ is sample-specific, the term 
${p_j}/{(1-p_j)}$
describes the effect of sample $j$ on read counts due to technical or biological reasons 
(\emph{e.g.}, different library sizes, biopsy sites, etc). 
We assume the relative expression of the $i$th gene in the $k$th subclone is described by $\phi_{ik}$, where $\phi_{ik}\geq 0$.  
Since the sample-specific effect has already been captured by $p_j$, for modeling convenience, we normalize the gene expression so that the expression levels sum to one for each subclone. Namely, $\sum_{i=1}^G\phi_{ik}=1$ for all $k=1,\cdots,K$. Furthermore, we assume that $\theta_{kj}$ represents the proportion of the $k$th subclone in the $j$th sample, where $\theta_{kj}\geq 0$ and $\sum_{k=1}^K\theta_{kj}=1$. We can interpret $\theta_{kj}\zeta_j$ as the population frequencies of the $k$th subclone in the $j$th sample, where parameter $\zeta_j$ controls the scale. 
Together, the summation $\sum_{k=1}^K\phi_{ik}\theta_{kj}\zeta_j$ represents the aggregated expression level of the $i$th gene across all $K$ subclones for the $j$th sample. 
To further account for 
the gene-specific random effects that are 
 independent of the samples and subclones, we introduce an additional term $\lambda\alpha_i$ to describe the random effect of the $i$th gene on the read counts such as GC-content and gene length. 
 We assume $\sum_{i=1}^G\alpha_i=1$ so that $\alpha_i$ represents the relative gene-specific random effect of the $i$th gene 
with respect to all the genes and $\lambda$ controls the overall scale of the gene-specific random effects.

Following \cite{zhou2016nonparametric}, the model in \eqref{NBFA_for_TH} has an augmented representation as
\begin{eqnarray}
y_{ij}&=&x_{ij}+z_{ij}, \nonumber\\
x_{ij}&=&\sum_{k=1}^Kx_{ijk}, \nonumber\\
\label{NBFA_random_effect1}
 z_{ij}\mid \lambda,\alpha_i,p_j&\sim&\mathrm{NB}(\lambda\alpha_i,p_j), \nonumber\\
\label{NBFA_random_effect2}
x_{ijk}\mid\bphi_{k},\btheta_j,\zeta_j, p_j&\sim&\text{NB}\left(\phi_{ik}\theta_{kj}\zeta_j,p_j\right).
\end{eqnarray}
From \eqref{NBFA_random_effect1}, 
the raw count $y_{ij}$ of the $i$th gene in the $j$th sample can be interpreted as coming from multiple sources: $x_{ijk}$ represents the count of the $i$th gene contributed by the $k$th subclone in the $j$th sample, where $k=1, \dots, K$, while $z_{ij}$ is the count contributed by the gene-specific random effect of the $i$th gene in the $j$th sample.


Denote $y_{\cdotv j} = \sum_{i=1}^Gy_{ij}$. Since $\sum_{i=1}^G \phi_{ik}=1$ and $\sum_{k=1}^K\theta_{kj} = 1$ by construction,  under \eqref{NBFA_random_effect1}, by the additive property of independent negative binomial random variables with the same 
success probability, we have
$$
y_{\cdotv j}\mid \lambda,\alpha_i,\zeta_j, p_j,\Phi,\Theta \sim\mathrm{NB}\left(\lambda+\zeta_j,~p_j\right),
$$
and, in particular, the mean as $\mathbb{E}[y_{\cdotv j}] = (\lambda + \zeta_j){p_j}/{(1-p_j)}$ and the variance as $\mathrm{Var}(y_{\cdotv j})= \mathbb{E}[y_{\cdotv j}] + \mathbb{E}^2[y_{\cdotv j}]/(\lambda+\zeta_j)$. 
It is clear that $p_j$, the between-sample random effect of the $j$th sample,   governs the variance-to-mean ratio of $y_{\cdotv j}$, whereas $\lambda+\zeta_j$, the sum of the scale $\lambda$ of 
the gene-specific random effects and the scale $\zeta_j$ for the $j$th sample,  controls the quadratic relationship between $\mathrm{Var}(y_{\cdotv j})$ and  $\mathbb{E}[y_{\cdotv j}]$.

We complete the model by setting the following priors that will be shown to be amenable to posterior inference: 
$$
\begin{aligned}
\bphi_k&\sim\mathrm{Dirichlet}(\eta,\cdots,\eta),\quad&&
\balpha\sim\mathrm{Dirichlet}(\delta,\cdots,\delta),\nonumber\\
\btheta_{j}\mid r_1,\cdots,r_K&\sim\mathrm{Dirichlet}\left(r_1,\cdots,r_K\right),\quad&& p_j\sim\mathrm{Beta}(a_0,b_0),\nonumber\\
\zeta_j\mid r_1,\cdots,r_K,c_j&\sim\mathrm{Gamma}\left(\sum_{k=1}^Kr_k,c_j^{-1}\right),\quad&&\lambda\sim\mathrm{Gamma}\left(u_0,v_0^{-1}\right),
\end{aligned}
$$
where $\bphi_k=(\phi_{1k},\cdots,\phi_{Gk})^T$, $\btheta_j=(\theta_{1j},\cdots,\theta_{Kj})^T$, $\balpha=(\alpha_1,\cdots,\alpha_G)^T$, $\mathrm{Gamma}(a, b)$ denotes a gamma distribution with mean $ab$ and variance $ab^2$, and $\mathrm{Dirichlet}(\eta_1,\cdots,\eta_d)$ denotes a $d$-dimensional Dirichlet distribution with parameter vector $(\eta_1,\cdots,\eta_d)$. We further impose the hyperpriors, expressed as $r_k\mid\gamma_0$, $c_0\sim\mathrm{Gamma}\left({\gamma_0}/{K},c_0^{-1}\right)$, $c_j\sim\mathrm{Gamma}\left(e_0,f_0^{-1}\right)$, $\gamma_0\sim\mathrm{Gamma}\left(g_0,h_0^{-1}\right)$, and $c_0\sim\mathrm{Gamma}\left(e_0,f_0^{-1}\right)$
to construct a more flexible model.



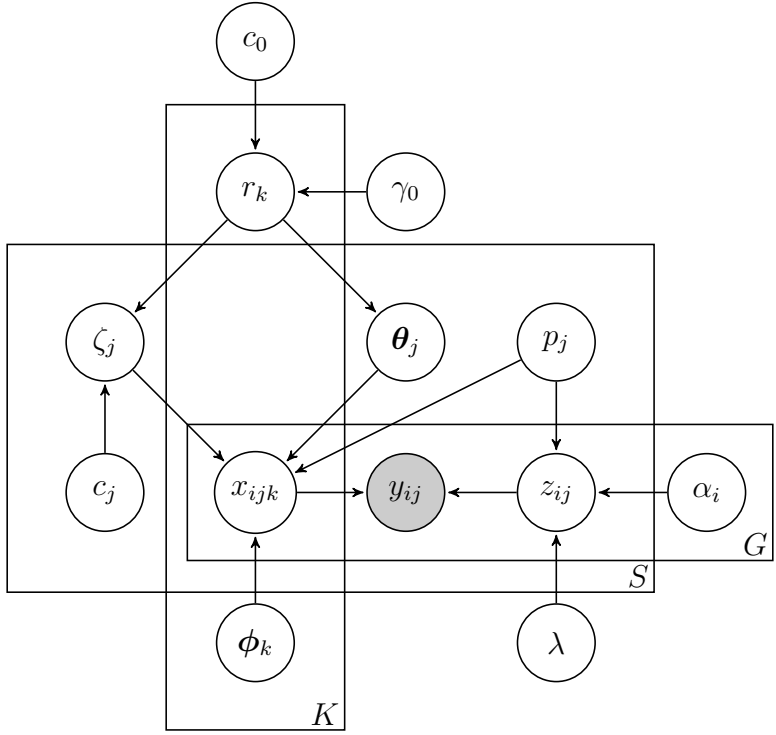
\begin{figure}[h]
	\centering
	\begin{tikzpicture}[->,>=stealth',shorten >=1pt,auto,node distance=2.8cm,semithick]
	\node[state]		(r) at (0, 2){$r_k$};
	\node[state]		(Theta) at (2,0){$\btheta_{j}$};
	\node[state]		(Phi) at (0,-4){$\boldsymbol{\phi}_k$};
	\node[state]		(x) at (0,-2) {$x_{ijk}$};
	\node[state]		(Z) at (4,-2) {$z_{ij}$};
	\node[state]		(lambda) at (4,-4) {$\lambda$};
	\node[state,fill=grau]		(Y) at (2,-2) {$y_{ij}$};
	\node[state]		(c) at (-2,-2) {$c_j$};
	\node[state]		(eta) at (-2,0) {$\zeta_j$};
	\node[state]		(p) at (4,0){$p_j$};
	\node[state]		(alpha) at (6,-2){$\alpha_i$};
	\node[state]		(gamma0) at (2,2){$\gamma_0$};
	\node[state]		(c0) at (0,4){$c_0$};
	\path
	(lambda) edge node {} (Z)
	(r) edge node {} (Theta)
	(c) edge node {} (eta)
	(r) edge node {} (eta)
	(eta) edge node {} (x)
	(p) edge node {} (x)
	(p) edge node {} (Z)
	(alpha) edge node {} (Z)
	(Theta) edge node {} (x)
	(Phi) edge node {}(x)
	(gamma0) edge node {}(r)
	(c0) edge node {}(r)
	(x) edge node {}(Y)
	(Z) edge node {} (Y);
	\node[draw, fit=(x)(c)(Theta)(p)(Z)(Y), inner sep=22pt]  (fit_x_c_Theta_p_Z_Y) {};
	\node[above left, inner sep=2pt] at (fit_x_c_Theta_p_Z_Y.south east) {$S$};
	\node[draw, fit=(x)(alpha)(Z)(Y), inner sep=10pt]  (fit_x_alpha_Z_Y) {};
	\node[above left, inner sep=2pt] at (fit_x_alpha_Z_Y.south east) {$G$};
	\node[draw, fit=(r)(x)(Phi), inner sep=18pt]  (fit_r_x_Phi) {};
	\node[above left, inner sep=2pt] at (fit_r_x_Phi.south east) {$K$};
	\end{tikzpicture}
	\caption{Graphical representation of the BayCount model. The boxes represent replicates. For example, 
	the box  containing $r_k,x_{ijk}$ and $\bphi_k$, with $K$ in its bottom right corner, 
	 indicates that there are $K$ ``copies" of $r_k,x_{ijk}$ and $\bphi_k$ with $k=1,\cdots,K$. Shaded nodes represent observations. }
	\label{NBFA_random_effect_graphical_model}
\end{figure}

Shown in Figure \ref{NBFA_random_effect_graphical_model} 
is  the 
graphical representation of our BayCount model. 

\subsection{Gibbs Sampling via Data Augmentation}
For the proposed BayCount model, while the conditional posteriors of $p_j$, $c_j$ and $c_0$ 
are straightforward to derive due to conjugacy, 
a variety of data augmentation techniques are used to derive the closed-form Gibbs sampling update equations
  for all the other model parameters. 
 Rather than going into the details here, let us first assume that we have already sampled the latent counts $x_{ijk}$ given the observations $y_{ij}$ and model parameters, which, according to Theorem 1 of  \citet{zhou2016nonparametric}, can be realized by sampling from the Dirichlet-multinomial distribution; given $x_{ijk}$, we show how to derive the Gibbs sampling update equations for $\Phi$ and $\Theta$ via data augmentation; and we will describe in the Supplementary Material a compound Poisson based blocked Gibbs sampler that completely removes the need of sampling $x_{ijk}$.


\subsubsection*{Sampling $\Phi$ and $\Theta$}
We introduce an auxiliary variable $\ell_{ijk}$ that follows a Chinese restaurant table (CRT) distribution, denoted by $\ell_{ijk}\mid x_{ijk},\phi_{ik}\theta_{kj}\zeta_j\sim \mathrm{CRT}(x_{ijk}, \phi_{ik}\theta_{kj}\zeta_j)$, with probability mass function
$$
 p(\ell_{ijk}\mid x_{ijk},\phi_{ik}\theta_{kj}\zeta_j) = \frac{\Gamma(\phi_{ik}\theta_{kj}\zeta_j)}{\Gamma\left({x_{ijk}+\phi_{ik}\theta_{kj}}\zeta_j\right)}|s(x_{ijk}, \ell_{ijk})| \left(\phi_{ik}\theta_{kj}\zeta_j\right)^{\ell_{ijk}}, 
$$
supported on $\{0,1,2,\cdots,x_{ijk}\}$,
where 
$s(x_{ijk}, \ell_{ijk})$ are Stirling numbers of the first kind \citep{johnson1997discrete}. Sampling $\ell\sim \mathrm{CRT}(x,r)$ 
can be realized by 
taking the summation of $m$ independent Bernoulli random variables: $
\ell=\sum_{t=1}^x b_t$, where $b_t\sim\mathrm{Bernoulli}\left(r/(r+t-1)\right)$ independently. Following \cite{zhou2012augment}, the joint distribution of $\ell_{ij}$ and $x_{ij}$ described by 
\begin{eqnarray}
\ell_{ijk}\mid x_{ijk} ,\phi_{ik},\theta_{kj},\zeta_j
&\sim&\mathrm{CRT}\left(x_{ijk},\phi_{ik}\theta_{kj}\zeta_j\right),\nonumber\\
x_{ijk} \mid\phi_{ik},\theta_{kj},\zeta_j,p_j
&\sim&\mathrm{NB}\left(\phi_{ik}\theta_{kj}\zeta_j,p_j\right)\nonumber,
\end{eqnarray}
can be equivalently characterized under the compound Poisson representation
\begin{eqnarray}
x_{ijk}\mid \ell_{ijk} ,p_j
&\sim&\mathrm{SumLog}\left(\ell_{ijk},p_j\right),\nonumber\\
\ell_{ijk} \mid\phi_{ik},\theta_{kj},\zeta_j, p_j
&\sim&\mathrm{Pois}\left(-\phi_{ik}\theta_{kj}\zeta_j\log(1-p_j)\right)\nonumber,
\end{eqnarray}
where $x\sim\mathrm{SumLog}\left(\ell,p\right)$ denotes the sum-logarithmic distribution generated as 
$x = \sum_{t=1}^{\ell}u_t$, where $(u_t)_{t=1}^\ell$ are independent, and identically distributed (i.i.d.) according to the logarithmic distribution \citep{quenouille1949relation} with probability mass function $p(u)=-p^{u}/[u\log(1-p)]$, supported on $\{1,2,\cdots\}$. 

Under this augmentation, the likelihood of $\phi_{ik}$, $\theta_{kj}$ and $\zeta_j$ becomes 
\begin{eqnarray}
\mathcal{L}(\phi_{ik},\theta_{kj},\zeta_j)\propto\Poi\left(\ell_{ijk}\mid-\phi_{ik}\theta_{kj}\zeta_j\log(1-p_j)\right),\nonumber
\end{eqnarray}
where $\Poi(\cdot\mid\lambda)$ denotes the probability mass function of the Poisson distribution with mean $\lambda$ \xx.  It follows immediately 
 that the full conditional posterior distributions for $\bphi_k$ and $\btheta_{j}$ are
\begin{eqnarray}
(\bphi_k\mid-)&\sim&\mathrm{Dirichlet}\left(\eta+
\sum_{j=1}^S\ell_{1jk},\cdots,\eta+\sum_{j=1}^S\ell_{Gjk}
\right),\nonumber\\
(\btheta_{j}\mid-)&\sim&\mathrm{Dirichlet}\left(
r_1+\sum_{i=1}^G\ell_{ij1},\cdots,r_K+\sum_{i=1}^G\ell_{ijK}
\right).\nonumber
\end{eqnarray}

Using data augmentation, we can similarly  derive the full conditional posterior distributions for $\zeta_j$, $\balpha$, $r_k$ and $\gamma_0$, as described in detail in the Supplementary Material.

\subsection{Determining the Number of Subclones $K$}
We have so far assumed \emph{a priori} that $K$ is fixed. Determining the number of factors in factor analysis 
is, in general, challenging. \citet{zhou2016nonparametric} suggested adaptively truncating  $K$ during Gibbs sampling iterations. 
This adaptive truncation procedure, which is designed to fit the data well, may tend to choose a large number of factors, some of which may be highly correlated to each other and hence 
appear to be redundant. 
To facilitate the interpretation of the model output,
we seek a model selection procedure that estimates $K$ in a more conservative manner. 
To select a moderate $K$ that is large enough to fit the data reasonably well, but at the same time is small enough for the sake of interpretation, 
we generalize the deviance reduction-based approach in \cite{shen2008forecasting} and calculate the estimated log-likelihood of the model under different numbers of subclones using post-burn-in MCMC samples. These samples are obtained by running the compound Poisson based blocked Gibbs sampler for different $K$'s. 
The estimate of $K$ can be  identified by an apparent decrease in the slopes of segments that connect the log-likelihood values of two consecutive $K$ values. Formally, we denote the log-likelihood function $\log\mathcal{L}(K)$ as a function of $K$, and define the second-order finite difference $\Delta^2\log\mathcal{L}(K)$ of the log-likelihood function by
$ \Delta^2\log\mathcal{L}(K):=2\log\mathcal{L}(K) -\log\mathcal{L}(K-1) - \log\mathcal{L}(K+1),$
 for $K=K_{\min}+1,\cdots,K_{\max}-1$, where $K_{\min}$ and $K_{\max}$ are the lower and upper limits of $K$, respectively.  
 Then an estimate of $K$ is given by 
$$\hat{K} = \mathrm{arg}\max_K\Delta^2\log\mathcal{L}(K).$$ 

\section{Simulation Study}\label{sec:simu}
In this section, we evaluate the proposed BayCount model through simulation studies.  Two different scenarios are considered. 

\begin{itemize}
	\item \textbf{Scenario I: } We simulate the data according to the BayCount model itself in \eqref{NBFA_for_TH}.  In particular, we generate the subclone-specific 
	gene expression data matrix $\Phi = (\phi_{ik})_{G\times K^o}\in\mathbb{R}_+^{G\times K^o}$ by i.i.d. draws of $\bphi_{k}\sim\mathrm{Dirichlet}(0.05,\cdots,0.05)$, 
	the proportion matrix $\Theta = (\theta_{kj})_{K^o\times S}$ by i.i.d. draws of $\btheta_j\sim\mathrm{Dirichlet}(0.5,\cdots,0.5)$, and $\zeta_j$ by i.i.d. draws of $\zeta_j\sim \mathrm{Gamma}(0.5K^o,1)$, where $i=1, \cdots, G$, $ j=1, \cdots, S$, and $k=1, \cdots, K^o$. 
Here  $G$ is the number of genes, $S$ is the number of samples, and $K^o$ is the simulated number of subclones.  
	We set $\lambda = 1$, draw $\balpha$ from $\mathrm{Dirichlet}(0.5,\cdots,0.5)$, and generate $p_j$ from a uniform distribution  such that the variance-to-mean ratio ${p_j}/{(1-p_j)}$ of $y_{\cdotv j}$ ranges from $100$ to $10^6$, encouraging the simulated data to be over-dispersed.  
	\item \textbf{Scenario II: }To evaluate the robustness of the BayCount model, under scenario II we consider simulating the data from a model that is different from the BayCount. 
	We generate the subclone-specific 
	gene expression data matrix $W = (w_{ik})_{G\times K^o}\in\mathbb{R}_+^{G\times K^o}$ by i.i.d. draws of $w_{ik}\sim\mathrm{Gamma}(0.05,10)$, 
	and the proportion matrix $\Theta = (\theta_{kj})_{K^o\times S}$ by i.i.d. draws of $\btheta_j\sim\mathrm{Dirichlet}(0.5,\cdots,0.5)$. 
	We set $\lambda=1$,  draw $\balpha$ from $\mathrm{Dirichlet}(0.5,\cdots,0.5)$, and generate $p_j$ from a uniform distribution such that the variance-to-mean ratio ${p_j}/{(1-p_j)}$ of $y_{\cdotv j}$ ranges from $100$ to $10^6$. 
	The count matrix $Y=(y_{ij})_{G\times S}$ is generated from 
	$
	y_{ij}\sim\mathrm{NB}\left(\lambda\alpha_i+\sum_{k=1}^{K^o}w_{ik}\theta_{kj},p_j\right)$. Note 
	that in scenario II the scales of $W=(w_{ik})_{G\times K^o}$ are not subject to the constraint $\sum_{i=1}^Gw_{ik}=1$. 
\end{itemize}


We will show that  BayCount  can accurately recover both the subclone-specific gene expression patterns and subclonal proportions. The hyperparameters are set to be $\eta = 0.1$, $a_0=b_0=0.01$, $e_0=f_0=1$, $g_0=h_0=1$, and $u_0=v_0=100$. We consider $K\in\{2,3,\cdots,10\}$. The compound Poisson based blocked Gibbs sampler is implemented with an initial burn-in of $B=1000$ iterations and a total of $n=2000$ iterations. The posterior means and  $95\%$ credible intervals for all parameters are computed using the $1000$ post-burn-in MCMC samples.

\subsection{Synthetic data with $K^o=3$}
\label{sub1}
We first simulate two 
datasets with $G=100$, $S=20$, and $K^o=3$ under both scenario I and scenario II. Under scenario I, the data generation scheme is the same as the BayCount model. Figure S1 in the Supplementary Material plots  $\Delta^2\log\mathcal{L}(K)$ versus $K$, indicating $\hat{K}=3$, which is the same as the simulation truth. The estimated subclone-specific gene expression matrix $\hat{\Phi}$ and subclonal proportions $\hat{\Theta}$ are computed as the posterior means of the post-burn-in MCMC samples. 
Figure S2 and S3 compare the simulated true $\Phi$ and $\Theta$ with the estimated $\hat{\Phi}$ and $\hat{\Theta}$, respectively. We can see that both the subclone-specific gene expression patterns and the subclonal proportions are successfully recovered. 

\begin{figure}[h!]
	\centerline{\includegraphics[width=0.8\textwidth]{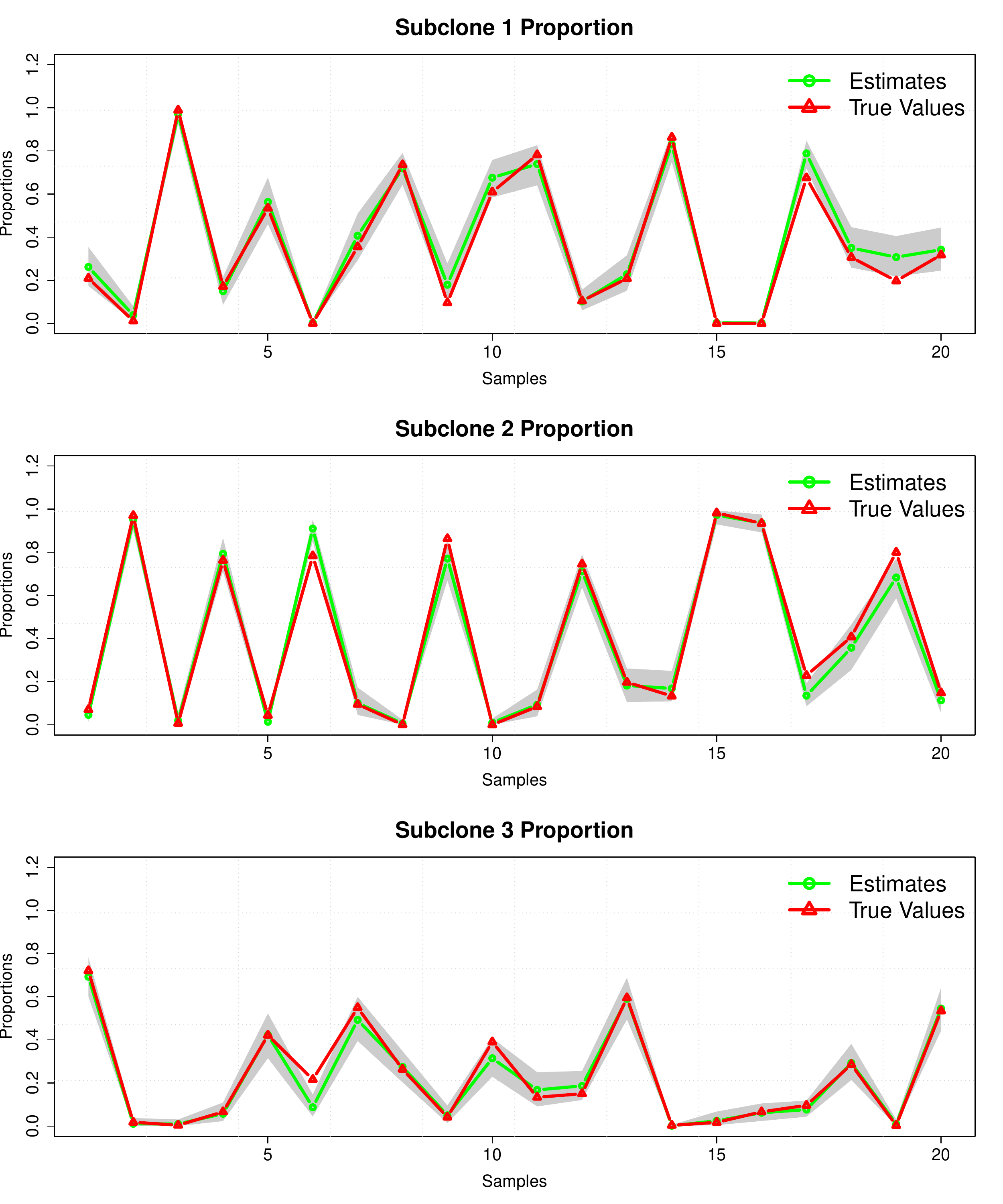}}
	\caption{The estimated subclonal proportions $\hat{\Theta}$ across samples $j=1,\cdots,20$ for the synthetic dataset with $K^o=3$ under scenario II. Horizontal axis is the index $j=1,\cdots,20$ of tumor samples, and vertical axis is the proportion. The green lines represent the estimate $\hat{\Theta}$, and the red lines represent the simulated true subclonal proportions. The shaded area represents the posterior $95\%$ credible bands. }
	\label{Subclones_Proportions_medium_scale_K}
\end{figure}

\begin{figure}[h!]
	\begin{center}
		\begin{tabular}{cc}
			\includegraphics[width=.5\textwidth]{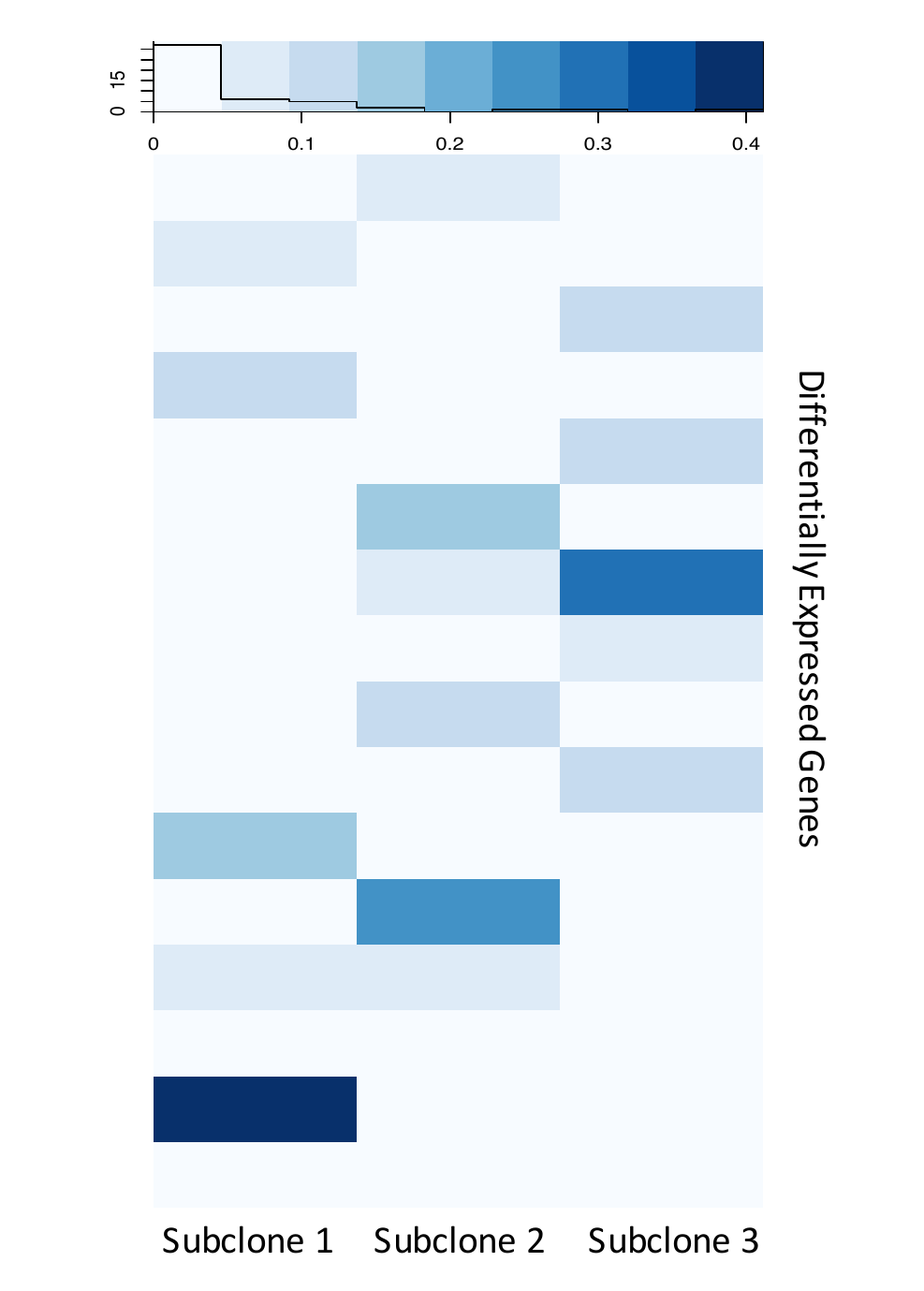}
			&
			\includegraphics[width=.5\textwidth]{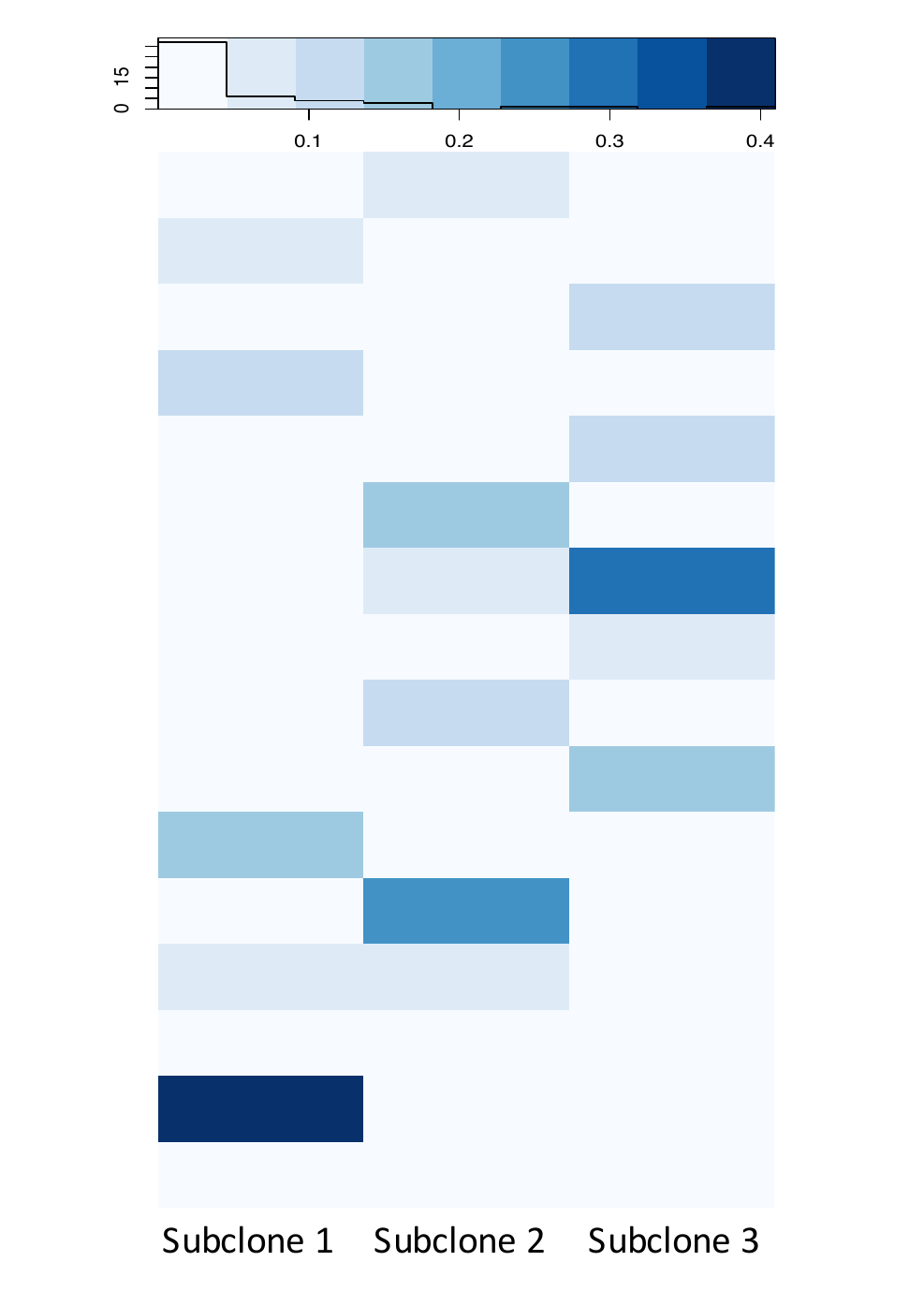}\\
			(a) $\hat{W}$& (b) $\hat{\Phi}$ \\
		\end{tabular}
	\end{center}
	\caption{Comparison of subclone-specific gene expression patterns 
	 for the synthetic dataset with $K^o=3$ under scenario II. Panel (a) is the heatmap of  $\hat{W}$, computed by normalizing the simulated true expression data $W$ by its column sums, and panel (b) is the heatmap of the estimate~$\hat{\Phi}$. }
	\label{Heat_map_gene_exp_K3}
\end{figure}

The analysis under scenario II is of greater interest, since the focus is to evaluate the robustness of BayCount. BayCount yields an estimate of $\hat{K}=3$, as shown in Figure S4. 
%
We then focus on the posterior inference based on $\hat{K}=3$. 
Figure \ref{Subclones_Proportions_medium_scale_K} compares the estimated subclonal proportions $\hat{\Theta}$ with the simulated true subclonal proportions across samples,
 along with the posterior $95\%$ credible intervals. The results show that the estimated $\hat{\Theta}$ approximates 
  the simulated true $\Theta$ well.
We then report the posterior inference on the subclone-specific gene expression $\Phi$. Under the BayCount model, $\sum_{i=1}^G\phi_{ik}=1$,  
hence the estimated $\hat{\Phi}$ by BayCount and the unnormalized gene expression profile matrix $W$ used in generating the simulated data  are not directly comparable. 
To see whether the gene expression pattern is recovered, we first normalize $W$ by its column sums as $\hat{W} = W\Lambda^{-1}$, where 
$
\Lambda=\mathrm{diag}\left(\sum_{i=1}^Gw_{i1},\cdots,\sum_{i=1}^Gw_{iK}\right)
$, so that $\hat{w}_{ik}$ represents the relative expression level of the $i$th gene in the $k$th subclone, and then compare $\hat{\Phi}$ with $\hat{W}$. 
For visualization, the genes with small standard deviations (less than $0.01$)
are filtered out due to their indistinguishable expressions across different subclones.  
Figure \ref{Heat_map_gene_exp_K3} compares the heatmap of $\hat{\Phi}$, with the heatmap of the simulated true (normalized) subclone-specific gene expression $\hat{W}$ on selected differentially expressed genes. 
It is clear 
 that the pattern of subclone-specific gene expression estimated by BayCount 
 closely matches the simulation truth.  

\subsection{Synthetic data with $K^o=5$}
Similarly as in Section \ref{sub1}, we simulate  two datasets with $G=1000$, $S=40$, and $K^o=5$ under scenarios I and  II, respectively. Under scenario I, BayCount yields an estimate of $\hat{K}=5$ (Figure S5), and from 
Figures S6 and S7, both the subclone-specific gene expression pattern and the subclonal proportions are successfully captured. 

\begin{figure}[htbp]
	\centerline{\includegraphics[width=0.9\textwidth]{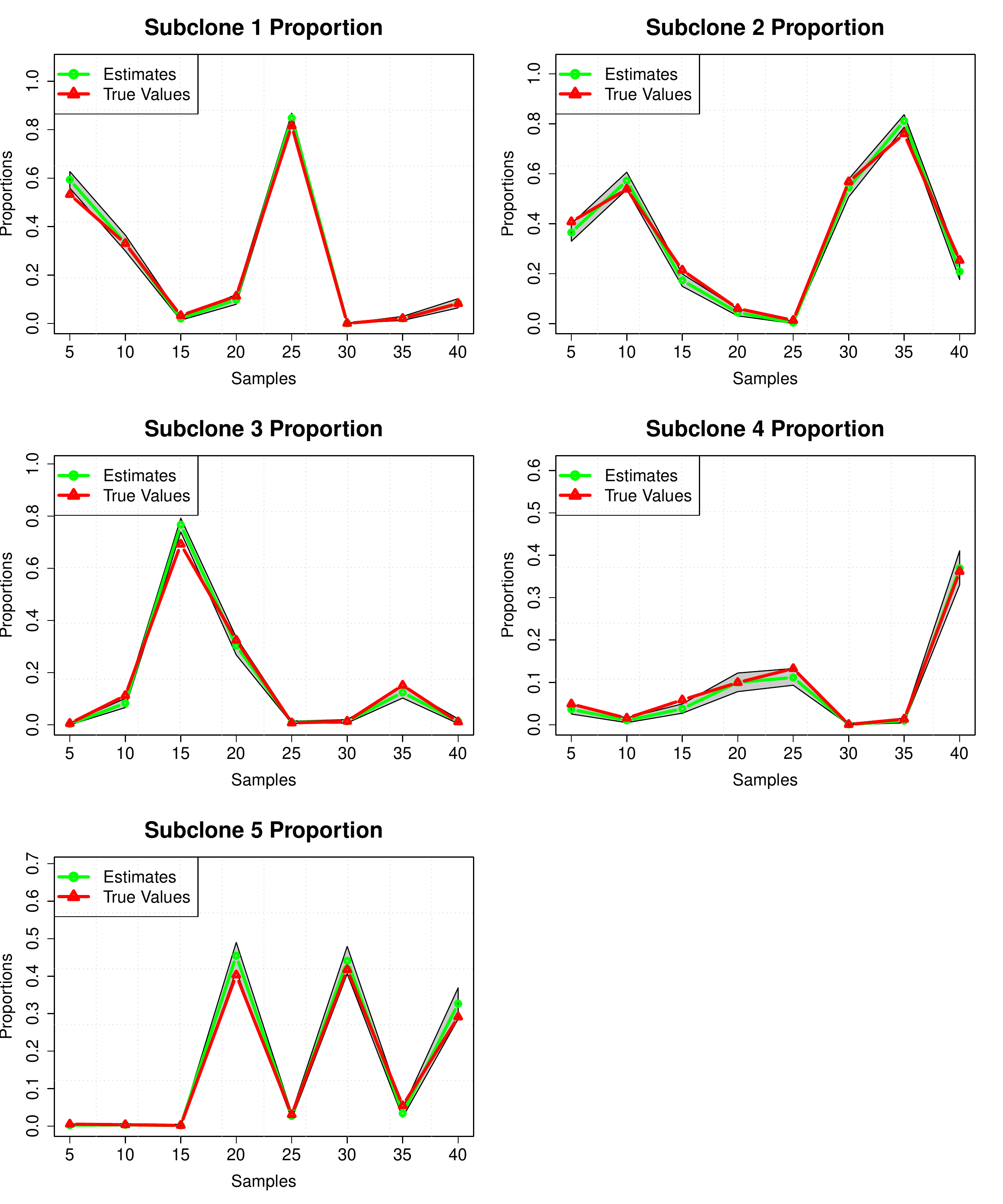}}
	\caption{Subclonal proportions across samples $j=5,10,\cdots,35,40$ for the synthetic dataset with $K^o=5$ under scenario II. Horizontal axis is the index 
	of tumor samples, and vertical axis is the proportion. The green lines represent $\hat{\Theta}$, and red lines represent the simulated true subclonal proportions. The shaded area represents the posterior $95\%$ credible bands. }
	\label{Subtypes_Proportions_medium_scale_K5_samples}
\end{figure}

Under scenario II, BayCount yields an estimate of $\hat{K}=5$ (Figure S8). 
For the subclonal proportions $\Theta=(\theta_{kj})_{K\times S}$, Figure  \ref{Subtypes_Proportions_medium_scale_K5_samples} shows that the estimated $\hat{\Theta}$ successfully recovers the simulated true proportions. Notice that the credible bands are narrower than those in Figure~\ref{Subclones_Proportions_medium_scale_K}, implying relatively smaller variability in estimating subclonal proportions for larger dataset. Figure S9
presents the autocorrelation plots of the posterior samples of some randomly selected proportions  by the compound Poisson based blocked Gibbs sampler, indicating  that the Markov chains mix well. 

\begin{figure}[h!]
	\begin{center}
		\begin{tabular}{cc}
			\includegraphics[width=.5\textwidth]{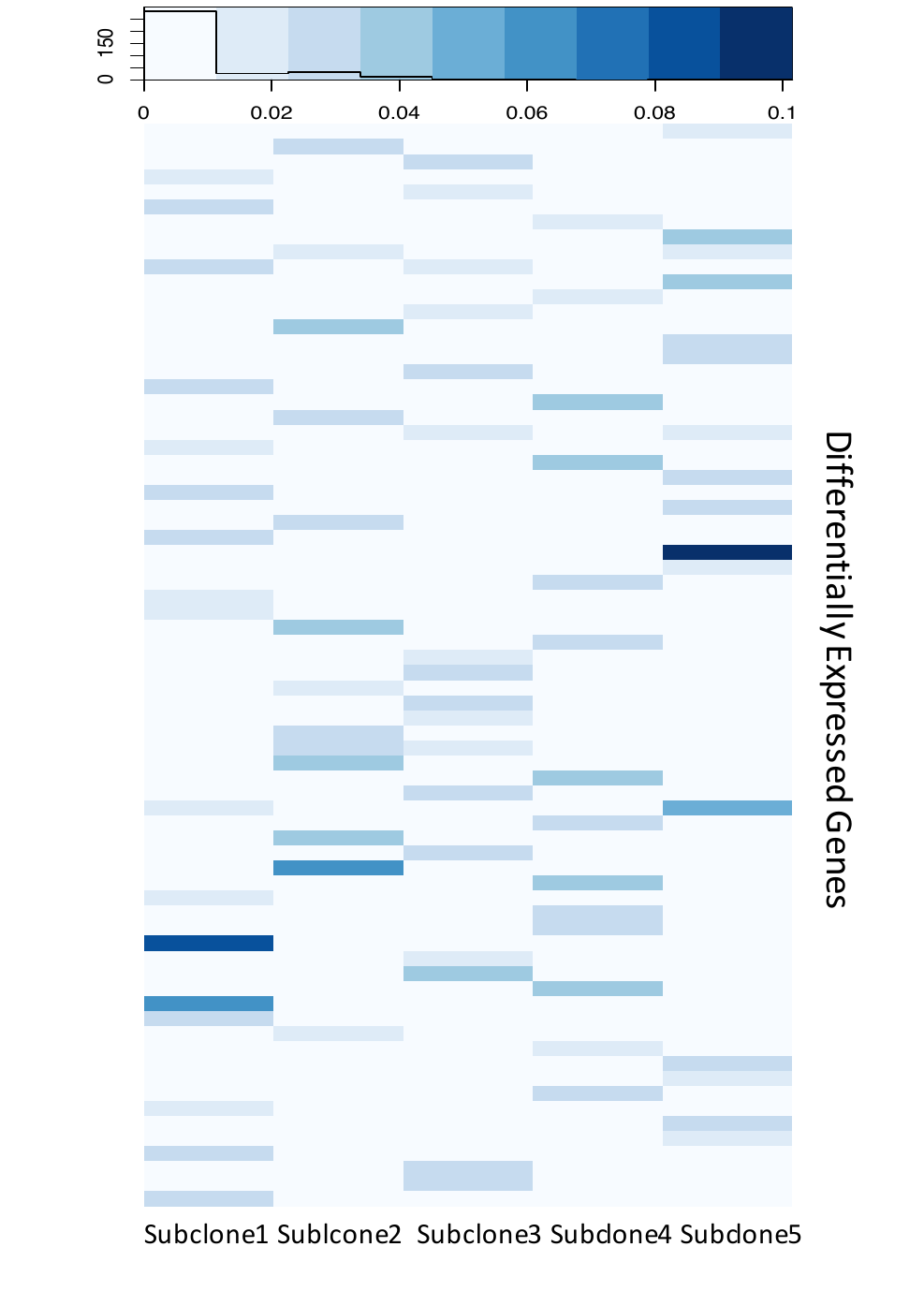}
			&
			\includegraphics[width=.5\textwidth]{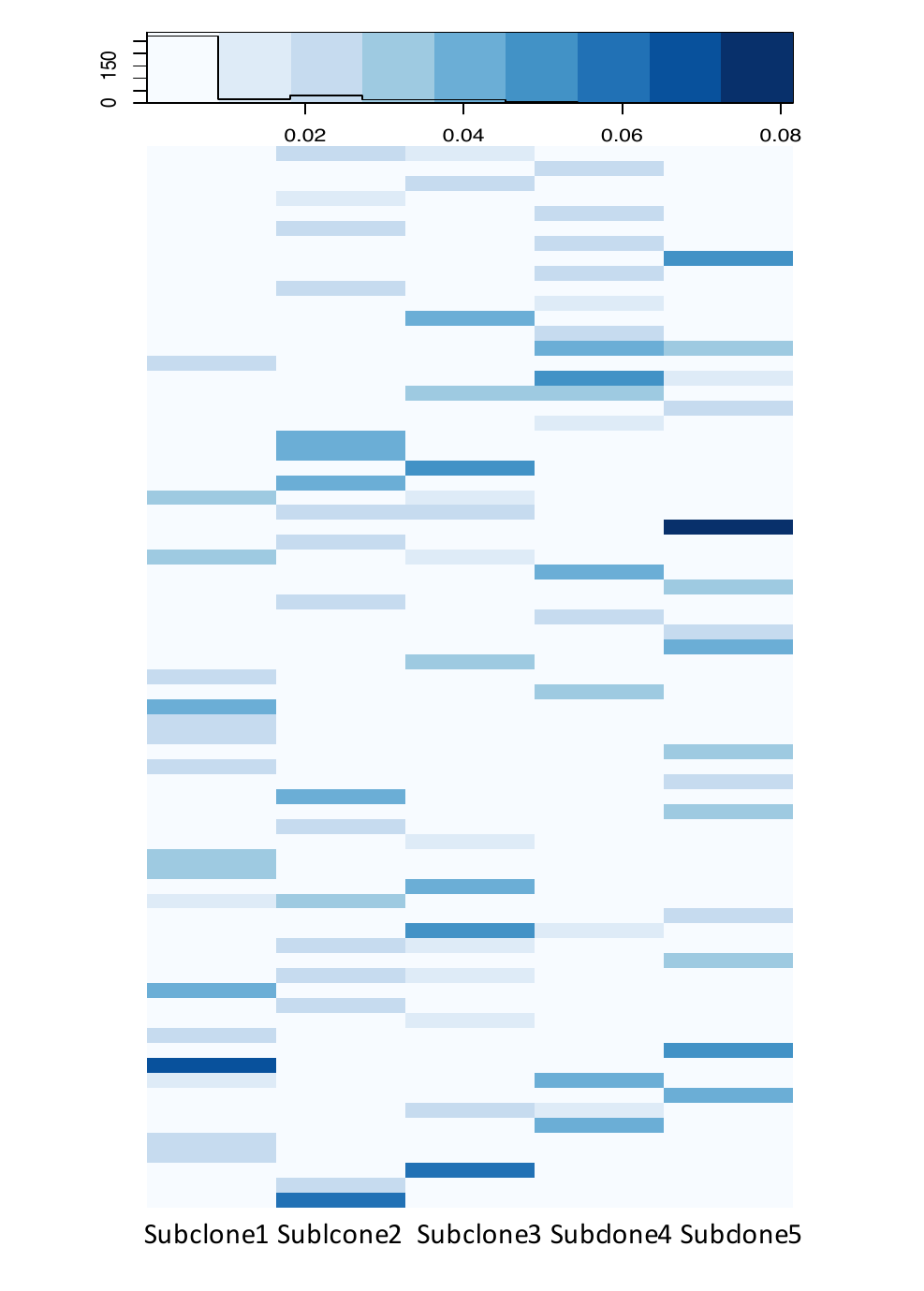}\\
			(a) $\hat{W}$& (b) $\hat{\Phi}$ \\
		\end{tabular}
	\end{center}
	\caption{Comparison of subclone-specific gene expression patterns 
	for the synthetic dataset with $K^o=5$ under scenario II. Panel (a) is the heatmap of  $\hat{W}$, computed by normalizing the simulated true expression data $W$ by its column sums, and panel (b) is the heatmap of the estimated $\hat{\Phi}$. }
	\label{Heat_map_gene_exp_K5}
\end{figure}

Figure \ref{Heat_map_gene_exp_K5} compares the simulated true (normalized) subclone-specific gene expression $\hat{W}$
with the estimated $\hat{\Theta}$ under the inference of BayCount. 
For this dataset we pre-screen $\hat{W}$ with a threshold $0.008$ on the across-subclone standard deviation for all genes for visualization. 
The high concordance between the heatmaps of the estimated and true expression patterns of the  differentially expressed genes indicates that the subclone-specific gene expression patterns have been successfully recovered as well.  


In summary, the BayCount model can accurately identify the number of subclones, estimate the subclonal proportions in each sample, and recover the subclone-specific gene expression pattern of the differentially expressed genes. 

\section{Real-world Data Analysis}
\label{sec:real_data}
We implement and evaluate the proposed BayCount model on the RNA-Seq data from The Cancer Genome Atlas (TCGA) \citep{cancer2012comprehensive} to study tumor heterogeneity (TH) in both lung squamous cell carcinoma (LUSC) and kidney renal clear cell carcinoma (KIRC). We first run the proposed Gibbs sampler for each fixed $K\in\{2,3,\cdots,10\}$, compute both the posterior mean and  $95\%$ credible interval of the log-likelihood for each fixed $K$, and estimate $K$ by maximizing $\Delta^2\log\mathcal{L}(K)$ over $K$. Next, based on the estimated $\hat{K}$ and the posterior samples generated by the proposed Gibbs sampler, we estimate the proportions of the identified subclones in each tumor sample and 
the subclone-specific gene expression,  which in turn can be used for a variety of downstream analyses.

\subsection{TCGA LUSC Data Analysis}
We apply the proposed BayCount model to the TCGA RNA-Seq data in lung squamous cell carcinoma (LUSC), which is a common type of lung cancer that  causes nearly one million
deaths worldwide every year. The raw RNA-Seq data for 200 LUSC tumor samples were processed using the Subread algorithm in the Rsubread package \citep{liao2014featurecounts} to obtain gene level read counts \citep{rahman2015alternative}. We select 382 previously reported important lung cancer genes \citep{wilkerson2010lung} for analysis, such as KRAS, STK11, BRAF, and RIT1. 

\begin{figure}[htbp]
	\centerline{\includegraphics[width=.8\textwidth]{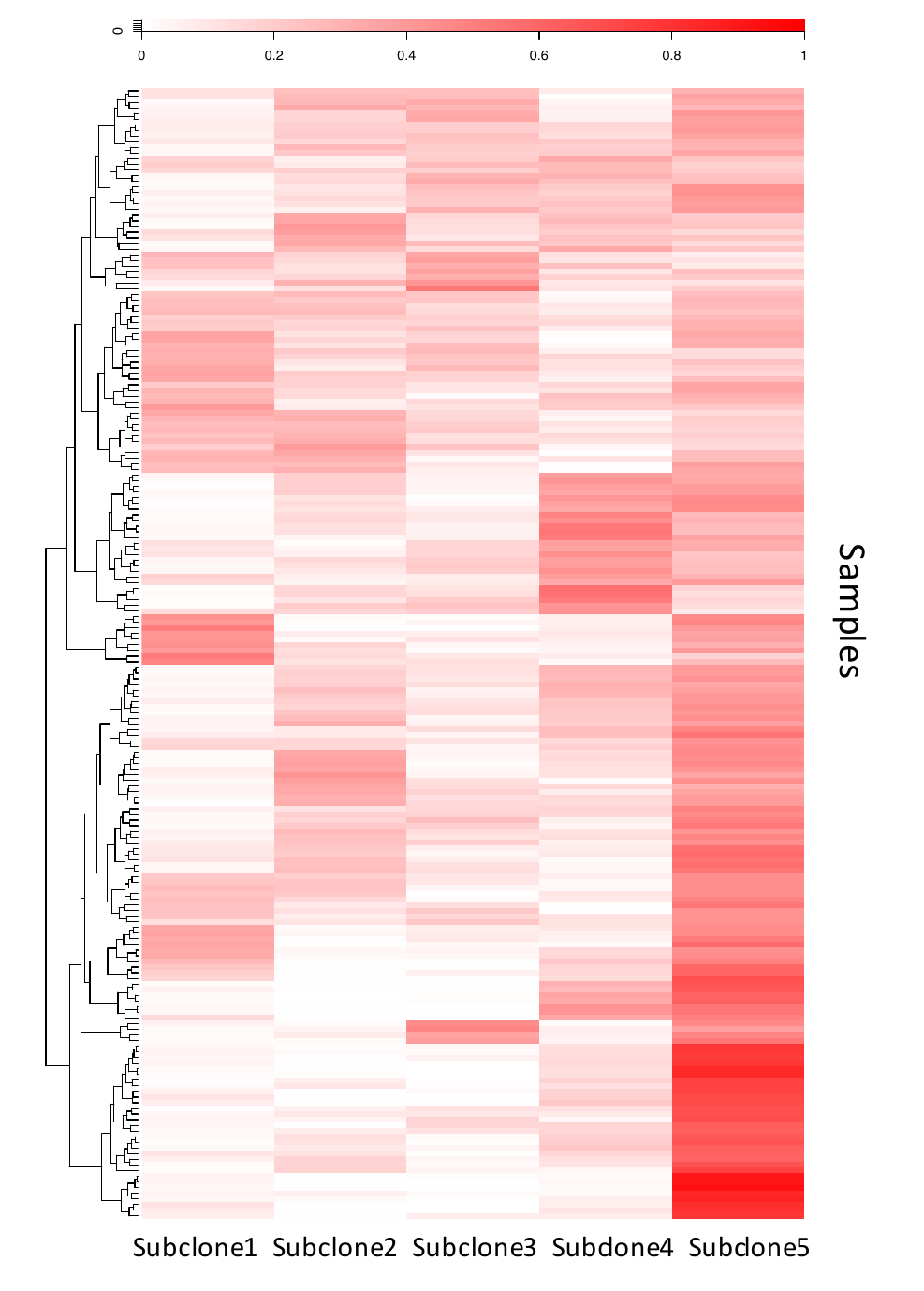}}
	\caption{Heatmap for the subclonal proportions across LUSC tumor samples $j=1,\cdots,200$. 
	From the heatmap it is clear that subclone 5 occupies relatively larger proportions for a large number of patients than the other 
	 4 subclones. }
	\label{lung_cancer_subclones_proportions_samples}	
\end{figure}

\begin{figure}[h!]
  \begin{center}
    \begin{tabular}{cc}
      \includegraphics[width=.6\textwidth]{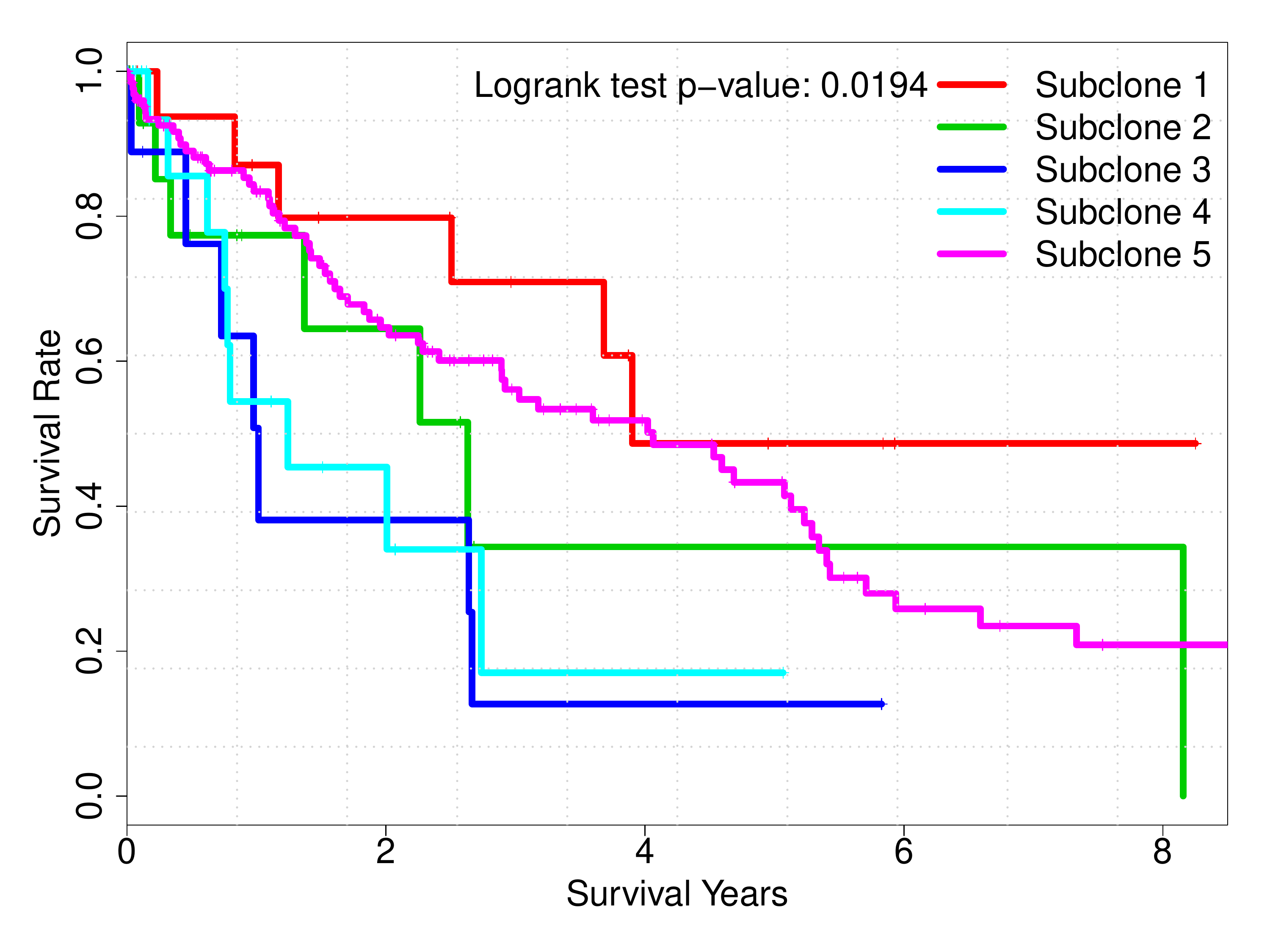}
      &
      \includegraphics[width=.4\textwidth]{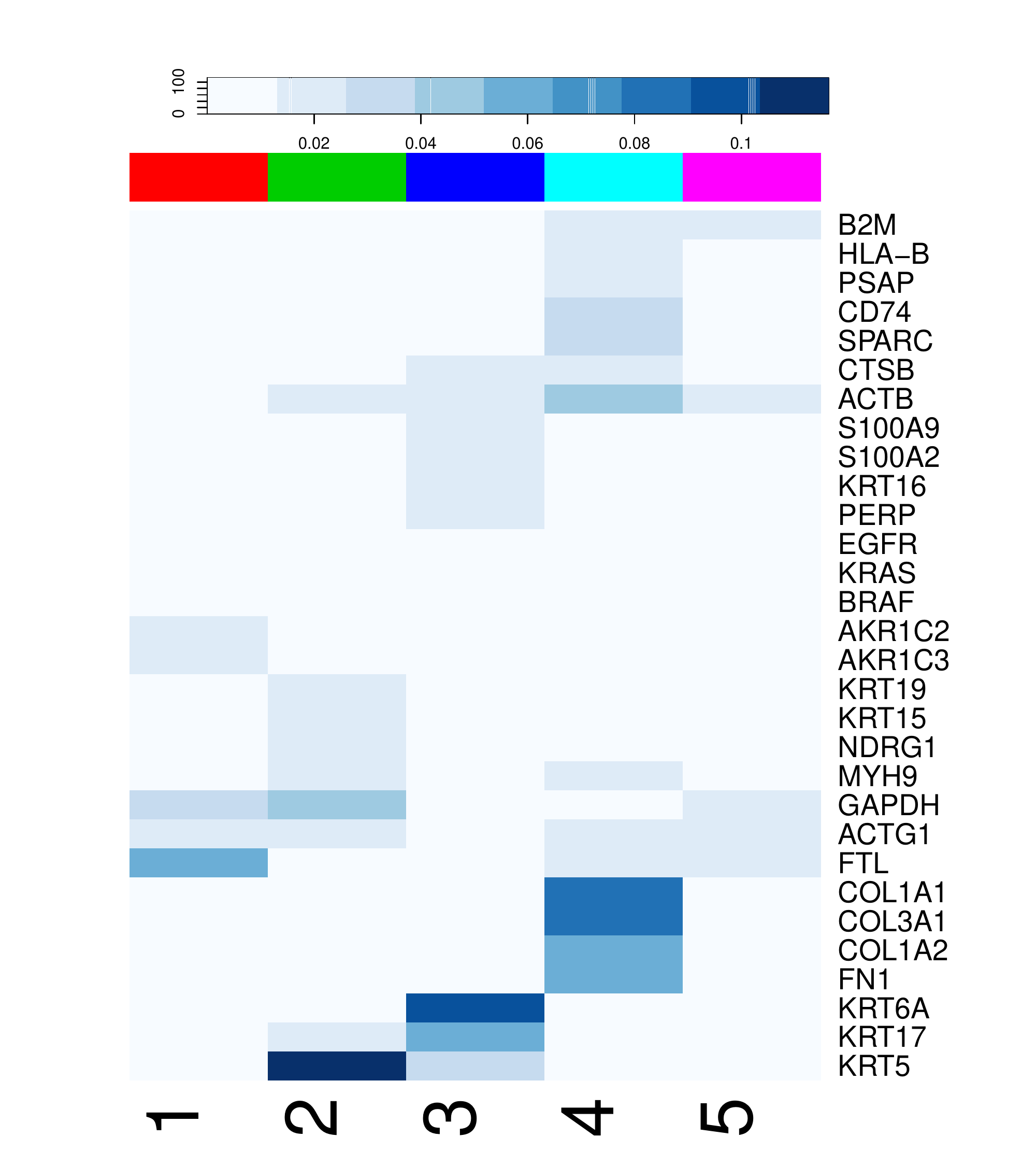}\\
      (a) & (b)  \\
    \end{tabular}
  \end{center}
	\caption{Panel (a) shows the Kaplan-Meier plots of overall survival in the LUSC dataset, where the patients are stratified by five clusters identified by subclone domination under the BayCount model. Panel (b) shows the subclone-specific gene expression of the top differentially expressed genes among five subclones.}
	\label{fig:lung}
\end{figure}

BayCount yields an estimate of five subclones (Figure S10) and their proportions in each tumor sample are shown in Figure \ref{lung_cancer_subclones_proportions_samples}. To identify the dominant subclone for each sample, we compare the estimated $\hat{\Theta}$ of the five subclones in each tumor sample, and use them to cluster the patients. Formally, for each patient $j=1,\cdots,S$, we compute the dominant subclone $
 k_j=\argmax_{k=1,\cdots,K} \hat{\theta}_{kj},$ and then cluster patients according to $\{j: k_j=k\}$, $k=1, \dots, \hat{K}$. 
That is to say, the patients with the same dominant subclone belong to the same cluster. We next check if the identified subclones have any clinical utility, \emph{e.g.}, stratification of patients in terms of overall survival.  Figure \ref{fig:lung}a shows the Kaplan-Meier plots of the overall survival of the patients in the five clusters identified by their dominant subclones. Indeed, patients stratified by these five BayCount-identified groups exhibit very distinct survival patterns (log-rank test $p = 0.0194$).
 
Figure \ref{fig:lung}b shows the expression levels of the top 30 differentially expressed genes  (ranked by the standard deviations of the  subclone-specific gene expression levels $\phi_{ik}$'s in an increasing order) in these five subclones. Distinct expression patterns are observed among  different subclones. For example, the FTL level is elevated in subclone 1; the expression levels of several genes encoding keratins (KRT5, KRT6A\zz, \xx etc.) are elevated in subclone 3; and the COL1A1 and COL1A2 expression levels are elevated in subclone 4. 
Interestingly, the patients with these dominant subclones also show the expected survival patterns. The subclone-1 dominated patients have better overall survival. Previous studies show 
 that the expression of FTL is decreased in lung tumors compared to normal tissues \citep{kudriavtseva2009expression}, and one plausible explanation is that subclone 1 may descend from less malignant cells and therefore resemble (or 
  consist  
  of) normal cells. Keratins and collagen I (encoded by COL1A1 and COL1A2) are known to play key roles in epithelial-to-mesenchymal transition (EMT), which subsequently initiates metastasis and promotes tumor progression \citep{depianto2010keratin, karantza2011keratins, shintani2008collagen}. 
 This agrees with our observation of worse prognosis in patients who have either subclone 3 (with elevated Keratin-coding genes) or subclone 4 (with elevated collagen I coding genes) as their dominant subclone.

\subsection{Kidney Cancer (KIRC) Data Analysis}

Similarly, we obtain gene level read counts \citep{liao2014featurecounts} for 200 TCGA kidney renal clear cell carcinoma (KIRC)  tumor RNA-seq samples and analyze them with BayCount. 
Among a total of 23,368 genes, 966 significantly mutated genes \citep{cancer2013comprehensive} in KIRC patients are selected, including VHL, PTEN, MTOR, etc. 

\begin{figure}[h!]
	\begin{center}
		\begin{tabular}{cc}
			\includegraphics[width=.6\textwidth]{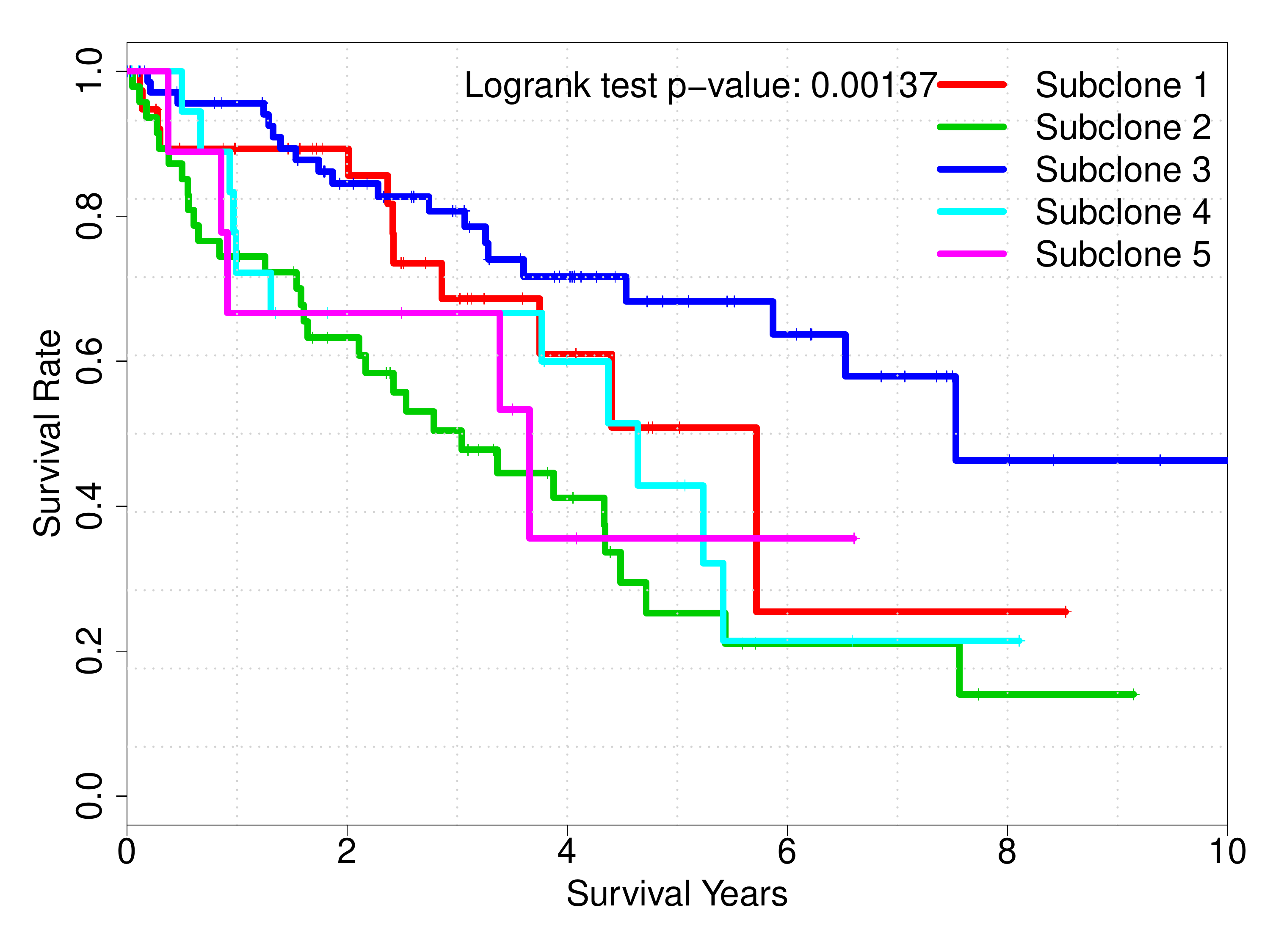}
			&
			\includegraphics[width=.4\textwidth]{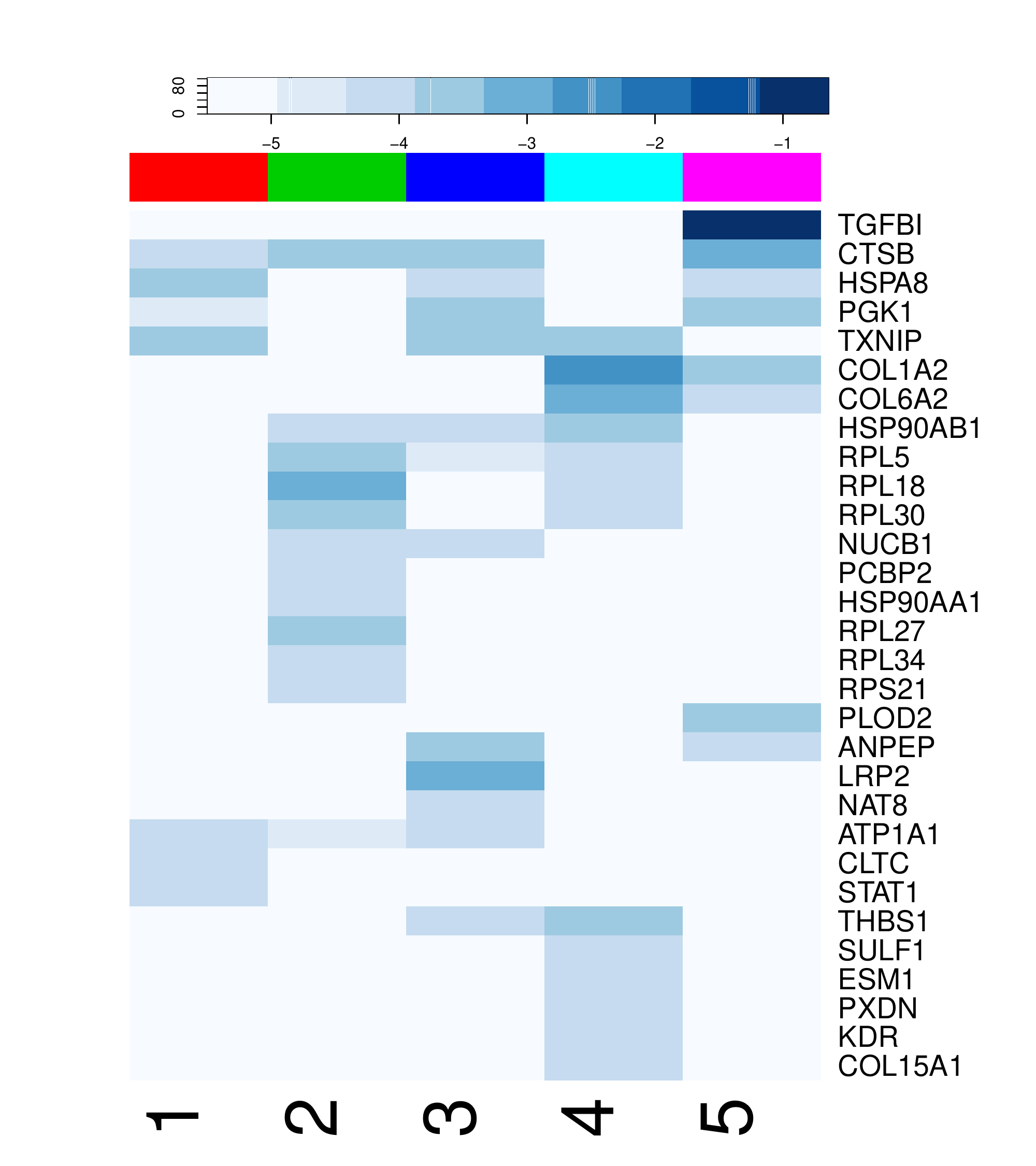}\\
			(a) & (b)  \\
		\end{tabular}
	\end{center}
	\caption{Panel (a) shows the Kaplan-Meier plots of overall survival in the KIRC dataset, where the patients are stratified by five clusters identified by subclone domination under the BayCount model. Panel (b) shows the subclone-specific gene expression (in the logarithmic scale) of the top differentially expressed genes among the five inferred  subclones.}
	\label{fig:kidney}
\end{figure}

BayCount yields an estimate of five subclones in KIRC (Figure S11). Figure \ref{fig:kidney} shows the Kaplan-Meier curves of the overall survival of the patients grouped by their dominant subclones (panel~a) and the heatmap of the top 30 differentially expressed genes (panel~b). Since we have a large number of genes to begin with, whereas $\sum_{i=1}^G\phi_{ik}=1$ for all $k=1,\cdots,K$, the subclone-specific gene expression estimates $\hat{\Phi}$ will be small. For better visualization, we plot $\hat{\Phi}$ in the logarithmic scale. 
The subclonal proportions across 200 KIRC tumor samples are shown in  Figure S12. 
As shown in Figure \ref{fig:kidney}, the patients with these dominant subclones again show distinct survival patterns. One of the poor survival groups (dominated by subclone 5) is characterized by elevated expression of TGFBI, which is known to be associated with poor prognosis \citep{zhu2015tgfbi} and 
matches our observation here. 

\begin{figure}[h!]
	\begin{center}
		\begin{tabular}{cc}
			\includegraphics[width=.5\textwidth]{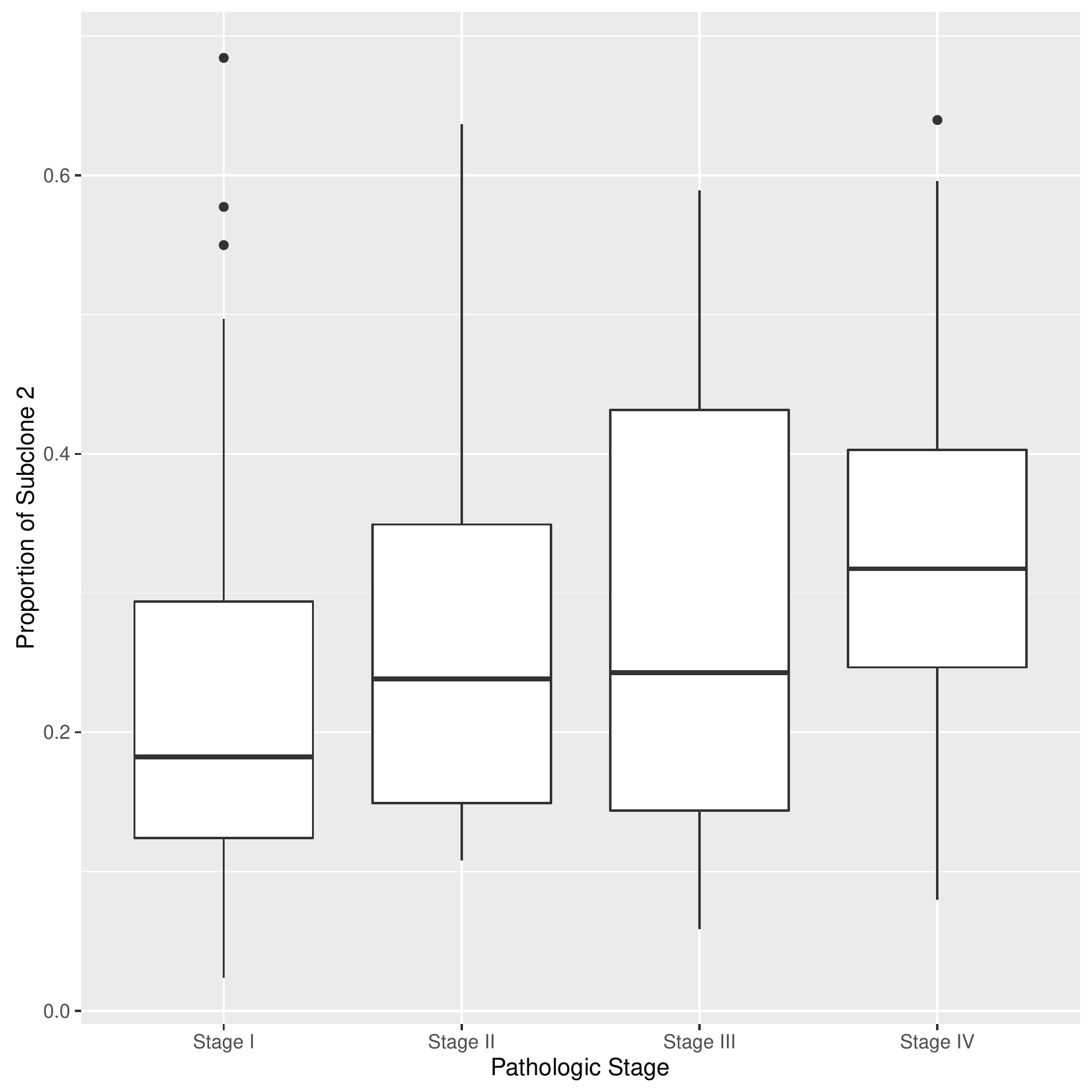}
			&
			\includegraphics[width=.5\textwidth]{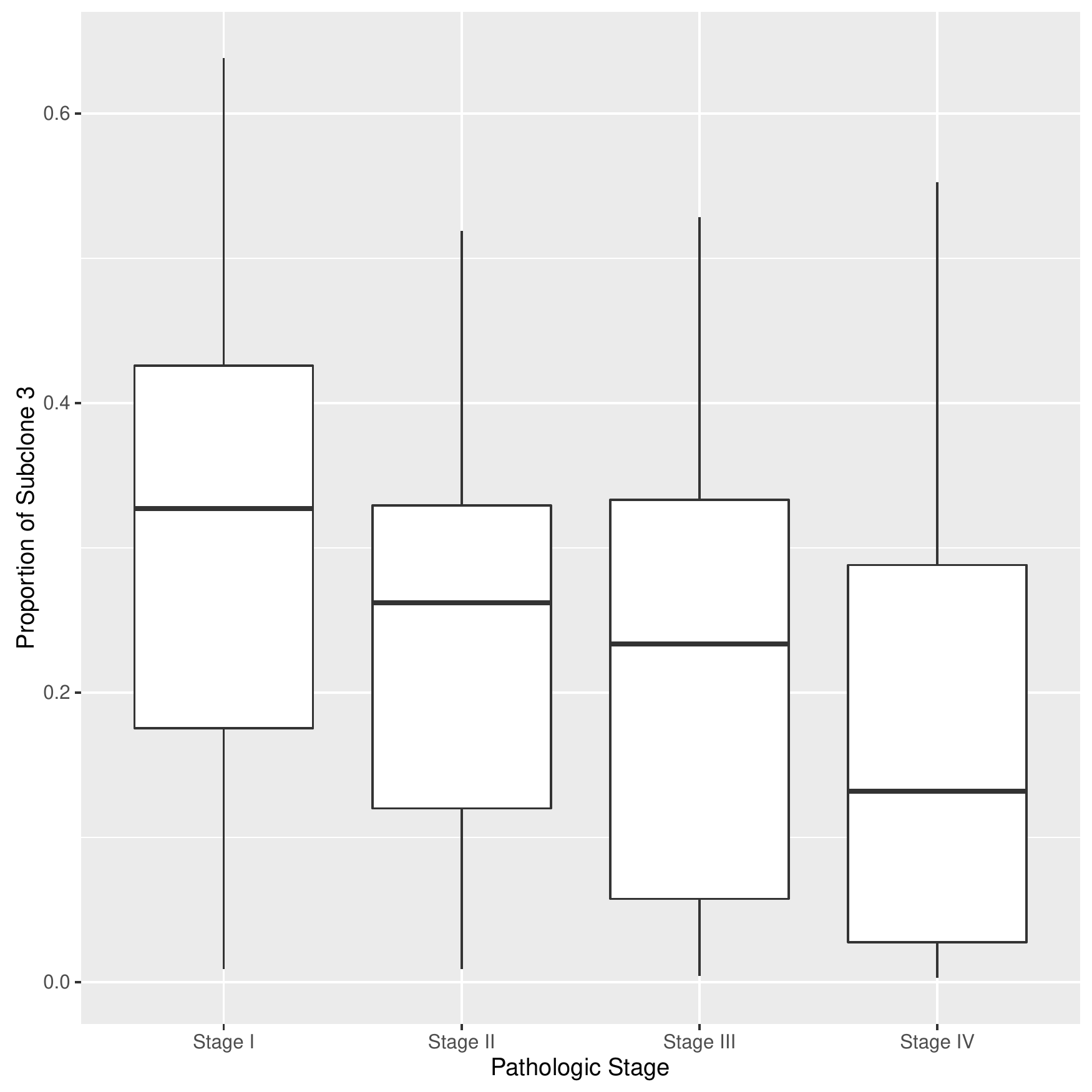}\\
			(a) Subclone 2& (b) Subclone 3 \\
		\end{tabular}
	\end{center}
	\caption{Panel (a): the proportions of subclone 2 in each tumor sample versus their pathologic stages ($p$-value $=0.00173$). Panel (b):  the proportions of subclone 3 in each tumor sample versus their pathologic stages ($p$-value $=0.00299$).}
	\label{fig:kidney_box}
\end{figure}

One distinction of our method from conventional subgroup analysis methods is that we focus on characterizing the underlying subclones (\emph{i.e.}, biologically meaningful subpopulations), by not only their individual molecular profiles but also their proportions. Instead of grouping the patients by their dominant subclones, we examine the proportion itself in terms of clinical utility. Interestingly, as shown in Figure \ref{fig:kidney_box}a, the proportion of subclone~2 increases with tumor stage: \emph{i.e.}, as subclone 2 expands and eventually outgrows other subclones, the tumor becomes more aggressive. In contrast, the proportion of subclone 3 decreases with tumor stage (Figure \ref{fig:kidney_box}b). Subclone 3 might be characterized by the less malignant (or normal-like) cells and takes more proportion in the beginning of the tumor life cycle. 
As tumor progresses to more advanced stages, subclone 3 could be 
suppressed by more aggressive subclones (\emph{e.g.}, subclone 2) and takes a decreasing proportion. Unsurprisingly,
the survival patterns 
 agree with our speculations about subclones 2 and 3, with 
the patients dominated by subclone 2 (the more aggressive subclone) and subcolone 3 (the less aggressive subclone) showing the worst and best survivals, respectively. 

More excitingly, we find that the proportions of these two subclones can complement clinical variables in further stratifying patients. For patients at early stage where the event rate is low and clinical information is relatively limited, the proportions of subclones 2 and 3 serve as a potent factor in further stratifying  patients (Figure S13) when dichotomizing at a natural cutoff. Combining our observations above, subclone proportions may provide additional insights into the progression course of tumors, assistance in biological interpretation, and potentially more accurate clinical prognosis.

\section{Conclusion} 
\label{sec:con}
The emerging high-throughput sequencing technology provides us with massive information for understanding tumors' complex microenvironment and allows us to develop novel statistical models for inferring tumor heterogeneity. Instead of 
normalizing RNA-Seq data that may bias 
downstream analysis, we propose BayCount 
 to directly analyze the raw RNA-Seq count data. Overcoming the natural challenges of analyzing raw RNA-seq count data, BayCount is able to factorize 
them while adjusting for both the between-sample and gene-specific 
 random effects.
Simulation studies show that BayCount can accurately recover the subclonal inference used to generate the simulated data. 
We apply BayCount to the TCGA LUSC and KIRC datasets, followed by correlating the subclonal inferences with their clinical utilities for comparison. In particular, by grouping patients according to their 
dominant subclones, we observe distinct and biologically sensible overall survival patterns for both LUSC and KIRC patients. Moreover, the proportions of the subclones may complement clinical variables in further stratifying patients.
In addition to prognosis value, tumor heterogeneity may be used as a biomarker to predict treatment response. 
For example, tumor samples with large proportions of cells bearing higher expressions on clinically actionable genes should be treated differently from those that have no or a small proportion of such cells. In addition, metastatic or recurrent tumors may possess very different compositions of subclones and should be treated differently. 


BayCount provides a general framework for inference on latent structures arising naturally in many other biomedical applications involving count data. For example, analyzing single-cell data is a potential further application of BayCount due to their sparsity and over-dispersion nature. 
\cite{macosko2015highly} describe Drop-Seq, a technology for profiling more than 40,000 single cells at one time. The unique characteristic of dropped-out events \citep{fan2016characterizing} in single cell sequencing limits the applicability of normalization methods in bulk RNA-Seq data. Also, 
such huge amount number of single-cells and high levels of sparsity pose difficulties for dimensionality reduction methods such as principal component analysis. 
Inferring distinct cell populations in single-cell RNA count data will be an interesting extension of BayCount.


\section*{Acknowledgement}
Yanxun Xu's research is partly supported by Johns Hopkins {\it in}Health and Booz Allen Hamilton. 

\bibliographystyle{apalike}
\bibliography{reference}

\end{document}